\documentclass[12pt,preprint]{aastex}
\usepackage{graphicx}
\shorttitle{A Double Main Sequence in NGC 6397} 
\shortauthors{A.\ P.\ Milone, et al.\ } 
\usepackage{ulem}

\begin{document}
\title{A Double Main Sequence in the Globular Cluster NGC 6397
          \footnote{           Based on observations with  the
                               NASA/ESA {\it Hubble Space Telescope},
                               obtained at  the Space Telescope Science
                               Institute,  which is operated by AURA, Inc.,
                               under NASA contract NAS 5-26555.}}
\author{
A.\ P. \,Milone\altaffilmark{2,3}, 
A.\ F. \,Marino\altaffilmark{4},
G.\ Piotto\altaffilmark{5, 6},
L.\ R. \,Bedin\altaffilmark{7},
J.\ Anderson\altaffilmark{7},  
A.\ Aparicio\altaffilmark{2,3}, 
S.\ Cassisi\altaffilmark{8}, 
R.\ M. \,Rich\altaffilmark{9}
 } 
\altaffiltext{2}{Instituto de Astrof\`\i sica de Canarias, E-38200 La
              Laguna, Tenerife, Canary Islands, Spain; [milone,aparicio]@iac.es}

\altaffiltext{3}{Department of Astrophysics, University of La Laguna,
           E-38200 La Laguna, Tenerife, Canary Islands, Spain}

\altaffiltext{4}{Max Plank Institute for Astrophysics, Postfach 1317,
	   D-85741 Garching, Germany; amarino@MPA-Garching.MPG.DE}

\altaffiltext{5}{Dipartimento  di   Astronomia,  Universit\`a  di Padova,
           Vicolo dell'Osservatorio 3, Padova I-35122, Italy;
           giampaolo.piotto@unipd.it }

\altaffiltext{6}{INAF-Osservatorio Astronomico di Padova, Vicolo             	 
	 dell'Osservatorio 5, I-35122 Padua, Italy}

\altaffiltext{7}{Space Telescope Science Institute,
                  3800 San Martin Drive, Baltimore,
                  MD 21218; [jayander,bedin]@stsci.edu \footnote{Since July 18th 2011 at INAF-Osservatorio Astronomico di Padova, Vicolo dell'Osservatorio 5, I-35122 Padova, Ital
y, EU.}}

\altaffiltext{8}{INAF-Osservatorio Astronomico di Collurania, via Mentore
           Maggini, I-64100 Teramo, Italy; cassisi@oa-teramo.inaf.it} 

\altaffiltext{9}{Division of Astronomy and Astrophysics, University of
  California, Los Angeles, 430 Portola Plaza, Box 951547, Los Angeles,
  CA 90095-1547, USA; rmr@astro.ucla.edu} 

\begin{abstract}
High-precision multi-band {\sl HST\/} photometry  reveals 
that the main sequence (MS) of the globular cluster NGC 6397 splits
into two components,
containing $\sim$30\% and $\sim$70\% of the stars. 
This double sequence is consistent with the idea that the cluster hosts
two stellar populations: ({\it i}) a primordial population that has a
composition similar to field stars, and containing $\sim$30\% of
the stars, and ({\it ii}) a second generation with
enhanced sodium and  nitrogen, depleted carbon and oxygen, and a
slightly enhanced helium abundance ($\Delta$Y$\sim$0.01).
We examine the color difference between the two sequences across a variety
of color baselines and find that the second sequence is anomalously faint in 
$m_{\rm F336W}$.  Theoretical isochrones indicate that this could be due 
to NH depletion.

\end{abstract}


%
%
%
\section{Introduction}
\label{introduction}
Globular clusters (GCs) were once thought to be composed of stars 
of a single composition and a single age, with few exceptions (such 
as $\omega$ Centauri).  It had long been known that most clusters 
contained red-giant stars with anomalies in their light-element 
abundances (Kraft 1979), but it was not clear whether this came as
a result of internal mixing or a variation in the primordial abundances.  

 High-resolution spectroscopy (e.\ g.\ Ramirez \& Cohen 2002,
  Carretta et al.\ 2009) has shown that these abundance anomalies in the giants 
     almost usually manifest themselves as an anticorrelation between Na 
     and O abundances, which is indicative of contamination from 
     high-temperture hydrogen-burning products (Denisenkov \& 
     Denisenkova 1989).   A similar anticorrelation has also been observed  
     in some un-evolved main-sequence (MS)  stars (Gratton et al.\ 2001, 
     Ramirez \& Cohen 2002), where the internal temperatures do not allow 
     hot CNO-cycle burning. This fact suggested the presence of two stellar 
     generations in the some cluster.

Indeed, high-precision photometry from {\it Hubble Space Telescope} 
({\it HST}) images has shown that $\omega$ Centauri and NGC 2808 host 
multiple distinct MSs (Anderson 1997, Bedin et al.\ 2004, Piotto et 
al.\ 2007), which have been associated with stellar populations with 
different helium (Norris 2004, D'Antona et al.\ 2005, Piotto et al.\ 2005), 
 with the bluer MSs having a higher He-content than the redder ones 
     (Norris 2004, Piotto et al.\ 2005,  D'Antona et al.\ 2005).
     More recently, a split MS has been observed also in 47 Tucanae 
     and NGC 6752 (Anderson et al.\ 2009, Milone et al.\ 2010, 2011a).  
     Stellar evolutionary models predict that H-burning at high 
     temperatures through the CNO cycle should result in enhanced 
     production of He.  In addition to He, such models also predict 
     enhanced production of N, and Na and depletion of C, and O.  This 
     pattern of enhancement/depletion has recently been confirmed among 
     the blue and red MS stars in NGC 2808 by Bragaglia et al.\ (2010).  

 Multiple populations have also been identified in the color-magnitude 
diagram (CMD) of some clusters in the form of  multiple sub-giant 
branches (SGBs, Milone et al.\ 2008, Marino et al.\ 2009) or multiple 
or anomalously  wide red-giant branches (RGBs, Marino et al.\ 2008, 
Yong et al.\ 2008, Lee et al.\ 2009).  
Multiple populations with discrete helium abundance may also offer 
an explanation for the complex HB morphology exhibited by some clusters 
(Busso et al.\ 2007, D'Antona \& Caloi 2008, Catelan, Valcarce \& 
Sweigart 2009).  A direct confirmation of a connection of the HB shape 
with the chemical content of HB stars comes from Marino et al.\ (2011), 
who have found that stars on the blue side of the instability strip of the 
cluster M4 are Na-rich and O-poor, whereas stars on the red HB are all 
Na-poor.

In this  paper, 
we present a study of multiple stellar populations in the nearby 
GC NGC 6397.  Evidence for a large spread in light-elements abundance, 
and a clear Na-O anticorrelation in this GC have been extensively 
documented in literature (Bell et al.\ 1979, Briley et al.\ 1990, 
Pasquini et al.\ 2004, Gratton et al.\ 2001, Carretta et al.\ 2005, 2009, 
Lind et al.\ 2009, 2011, L11).  Recently, Richer et al.\ (2006) used 
the Advanced Camera for Surveys (ACS) of the {\it HST} for 126 orbits 
to image an outer region of NGC 6397 in the F606W and F814W bands.   
The spectacular CMD they obtained reaches the deepest intrinsic 
luminosities for a GC achieved thus far, and reveals an extermely tight 
MS in the ${\it m}_{\rm F606W}-{\it m}_{\rm F814W}$ color system
(Anderson et al.\ 2008a).  These observations are consistent with negligible
He variation among NGC 6397 stars:  $\Delta$Y $<$ 0.02 (Di Criscienzo 
et al.\ 2009).  Here, we take advantage of the multi-band {\it HST} 
photometry of the central region to study the MS of NGC 6397 for signs 
of multiple populations.  The paper is organized as follows.  
 The data and the data reduction are described in Sect.~\ref{sec:data}, 
     results are shown in Sect.~\ref{sec:MS}, and discussed in 
     Sect.~\ref{sec:discussion}.  

%
%
%
\section{Observations and data reduction of NGC 6397} 
\label{sec:data}
\begin{figure}[ht!]
\centering
\epsscale{.52}
\plotone{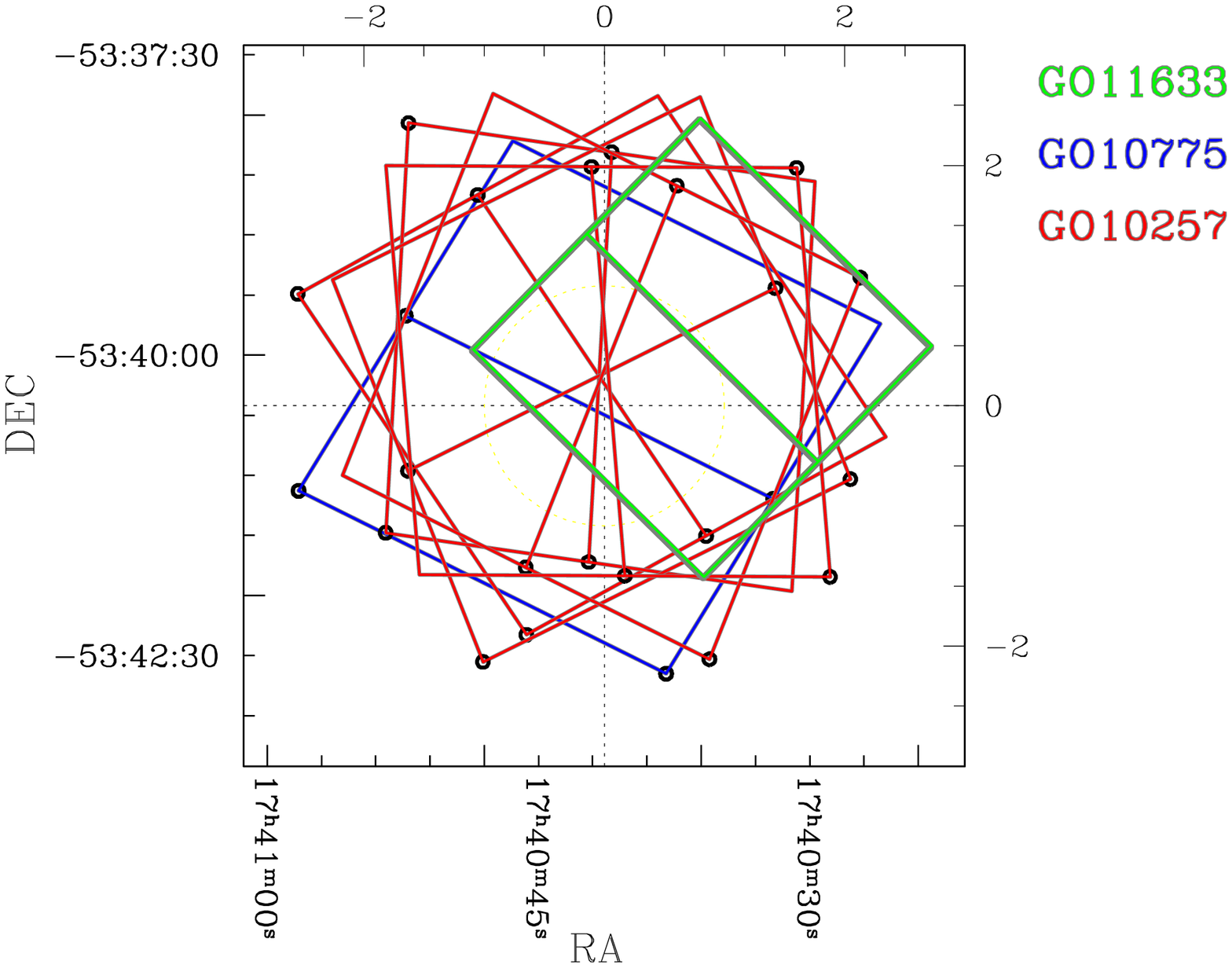}
\epsscale{.42}
\plotone{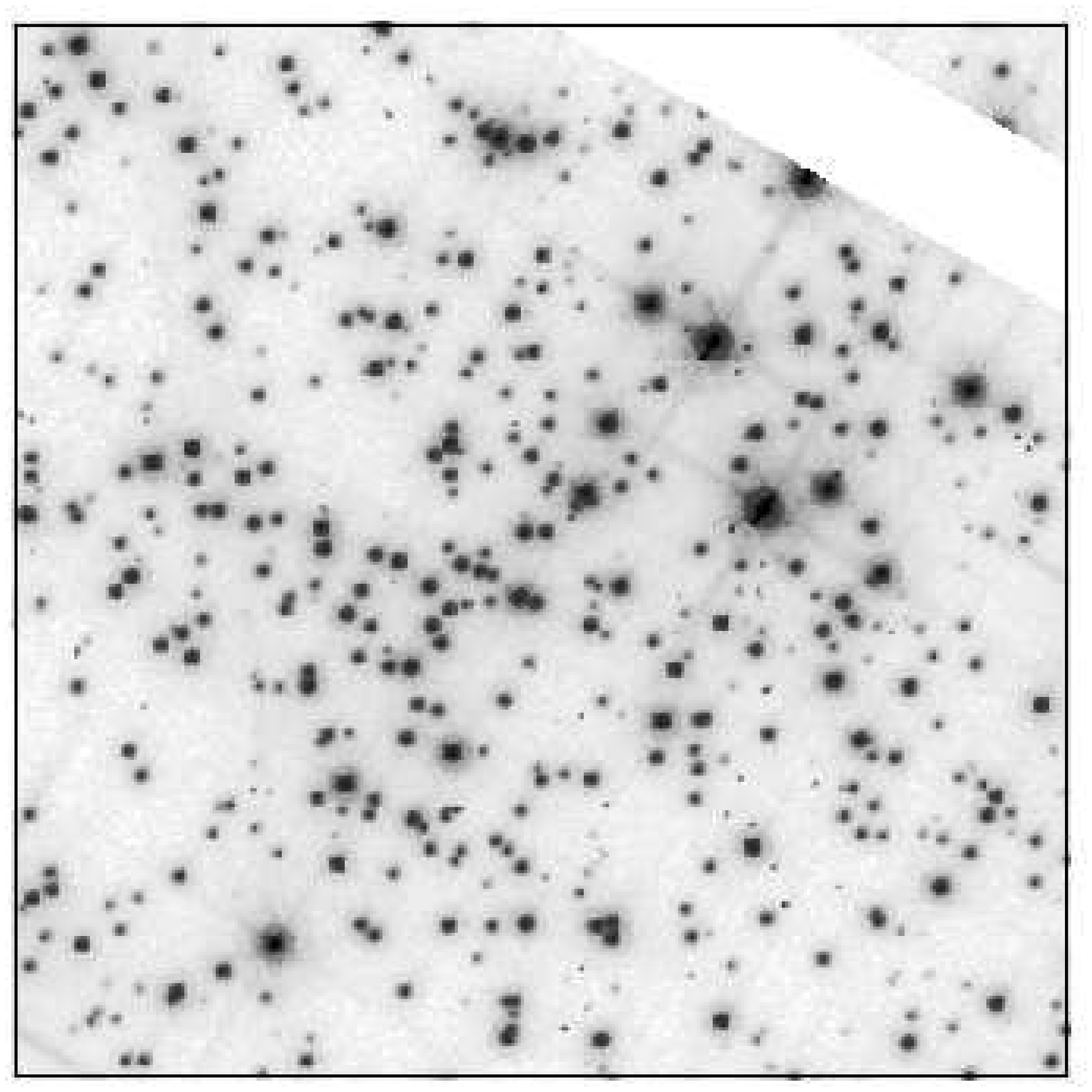}
\caption{ 
 {\textit Left panel:} Footprint of the {\it HST} data sets used in
   this paper. The small circles mark the corner of the WFC chip \#1.  
{\textit Right panel:} 10$\times$10 arcsec field of a single F814W image  
centered on the cluster center.
} 
         \label{FOOT}
   \end{figure}

For this project we used {\it HST} images taken at three different epochs 
with the Wide Field Channel of the ACS (ACS/WFC) 
and the ultraviolet/visible channel of the Wide-Field Camera 3 (WFC3/UVIS).
In particular, we used the following data sets (program-ID, camera, date, 
filters, number of exposures, exposure time): 
     (1) GO-11633, WFC3/UVIS, March 2010, 
                   F336W, 6$\times$620 s and 
                   F225W, 22$\times$680 s; 
     (2) GO-10257, ACS/WFC, August 2004-June 2005, 
                   F435W, 5$\times$13 s $+$ 5$\times$340 s; 
     (3) GO-10775, ACS/WFC, May 2006, 
                   F606W, 1 s $+$4$\times$15 s and 
                   F814W, 1 s $+$4$\times$15 s.  
The field covered by these images is centered on the cluster
core. 
 Their footprint are shown in Fig.~\ref{FOOT}, together with
        an image of the 10$\times$10 arcsec central field taken 
        through the F814W filter.  Clearly, crowding is not severe, 
        even in the center of this nearby cluster.

Star positions and fluxes in the ACS/WFC images were measured with the
software described by Anderson et al.\ (2008b).  The program analyzes 
a set of exposures simultaneously in order to produce a catalogue of 
stars over the field of view.  Stars are measured in each image independently 
by means of a spatially variable point-spread-function model from 
Anderson \& King (2006), plus a spatially constant perturbation of the 
PSF that accounts for the effects of focus variations.  The photometry 
has been calibrated as in Bedin et al.\ (2005) using the encircled energy 
and zero points of Sirianni et al.\ (2005).

The WFC3/UVIS images were reduced with a software that is adapted from
img2xym\_WFI (Anderson et al.\ 2006).  Astrometry and photometry were 
corrected for pixel area and geometric distortion as in Bellini, 
Anderson \& Bedin (2011).  The proper motions were derived as in 
Milone et al.\ (2006). 
 
\begin{figure}[ht!]
\centering
\epsscale{.55}
\plotone{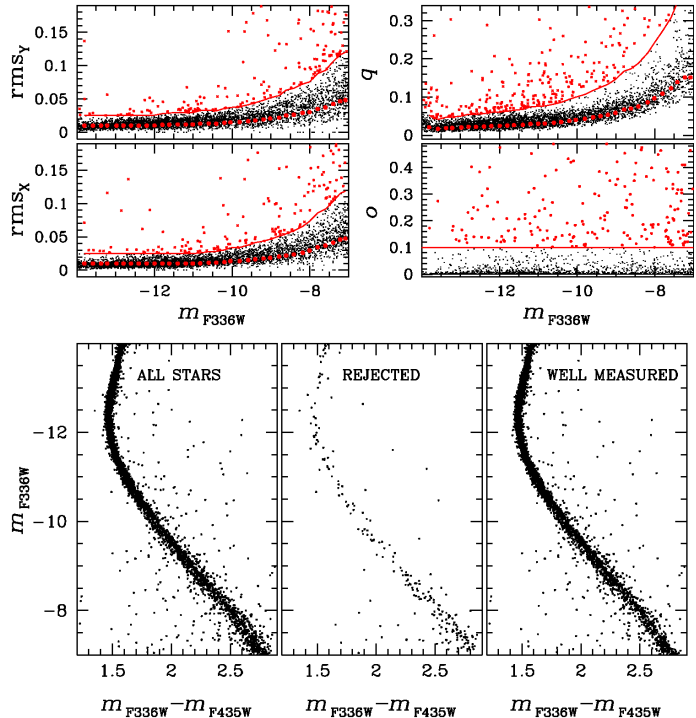}
\caption{ 
 \textit{Upper panels:}
 Diagnostic parameters used to select the stars with the best
 photometry are plotted as a function of $m_{\rm F336W}$. Red lines
 separate the well measured stars (thin black points) from those that
 are more likely to have a poorer photometry (red thick points). 
 \textit{Lower panels:}
 Comparison of the CMD of all the measured stars (left), of rejected
 stars (middle), and of stars that pass our criteria of selection (right).
} 
         \label{M1}
   \end{figure}

This paper is mainly based on high-precision photometry, and is limited 
to a sub-sample of stars that are relatively isolated, are well-fit by 
the PSF, and also have small photometric and astrometric errors.  
 The photometric software provides a number of quality indices
  that can be used as diagnostics of the reliability of photometric
  measurements. Specifically, these are: 
  (1) the rms of the positions measured in different exposures 
      trasformed in a common reference frame ($rms_{\rm X}$ and 
      $rms_{\rm Y}$ ), 
  (2) the residuals to the PSF fit for each star ($q$), and 
  (3) the ratio between the estimated flux of the star in a 0.5 arcsec 
      aperture and the flux from neighbouring stars within the same 
      aperture ($o$, see Anderson et al.\ 2008b for details.)
To select this 
 high-precision
sub-sample of stars, we 
followed the procedure described by Milone et al.\ (2009, Sect.~2.1) 
 and illustrated in the upper panels of Fig.~\ref{M1} for NGC 
     6397. As an example, in the lower panels we show the 
     $m_{\rm F336W}$ vs. $m_{\rm F336W}-m_{\rm F435W}$ CMD of all the 
     measured stars, of rejected stars, and of stars that pass the adopted 
     criteria of selection.  Since NGC 6397 ($l \simeq 338^{\circ}$, 
     $b \simeq -12^{\circ}$) is located in a Galactic field richly 
     populated by Bulge and disk stars, we must make every effort to 
     remove from our final list stars that do not belong to the cluster.


\begin{figure}[ht!]
\centering
\epsscale{.45}
\plotone{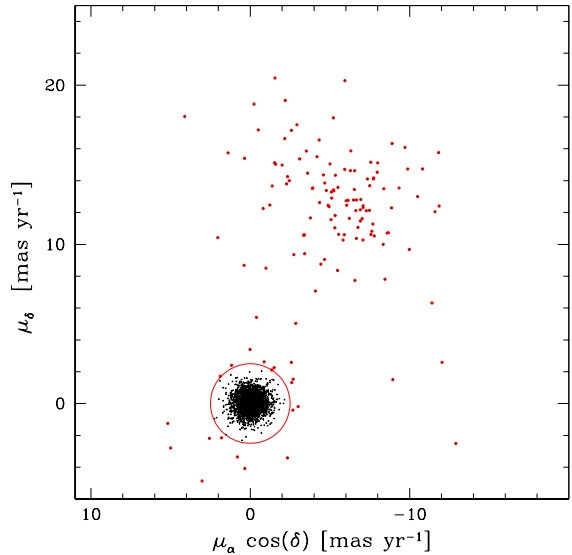}
\caption{ Vector-point diagram of the proper motions in equatorial
  coordinates. The red circle separates cluster members (black points)
  from field objects (red points).  
} 
         \label{M2}
   \end{figure}

 We measured proper motions using the F435W images of GO-10257
     data set and the GO-10775 images, collected in two epochs separated
     by $\sim$2 years.  Figure~\ref{M2} shows the vector-point diagram 
     of proper motions. The separation between field stars and cluster 
     stars is very well defined and cluster members were selected on 
     the basis of their common proper motion. 
     In all CMDs discussed in the following, field stars have been 
     removed using the proper motion selection criterion.

The average reddening for NGC 6397 is $E(B-V)=0.18$ (Harris 1996, 2010).
Usually, such a large reddening is not uniform 
  across an ACS field .  
A visual inspection at the CMDs of NGC 6397 indeed reveals that the 
sequences are 
broadened
by differential reddening.  We corrected the effect of differential 
reddening by means of a procedure that is described in detail in Milone 
et al.\ (2011b), and used in many other papers by these authors.

Briefly, we define the fiducial MS 
       ridge-line in a given color system
for the cluster and estimate, for each 
star, how the observed stars in its vicinity may systematically lie to 
the red or to the blue of the fiducial sequence; this systematic color 
and magnitude offset, measured along the reddening line, is indicative 
of the local differential reddening. 

\begin{figure}[ht!]
\centering
\epsscale{.6}
\plotone{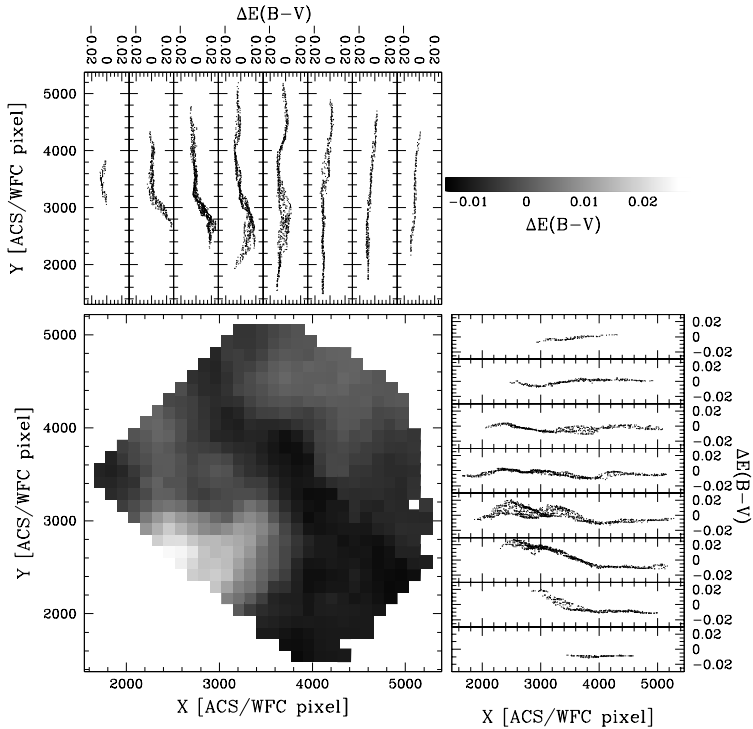}
\caption{ 
	\textit{Bottom-left}: Map  of  differential reddening in the
        NGC 6397 field of view.  The gray levels
        correspond to the magnitude  of the variation  in local
        reddening, as indicated in the   upper-right panel.
	We also divided the field of view into 8 horizontal slices
	and 8 vertical slices. \textit{Upper-left} and
        \textit{lower-right} panels show $\Delta~E(B-V)$ as a function
        of the Y and X coordinates in each slide.  } 
         \label{M3b}
   \end{figure}

\begin{figure}[ht!]
\centering
\epsscale{.55}
\plotone{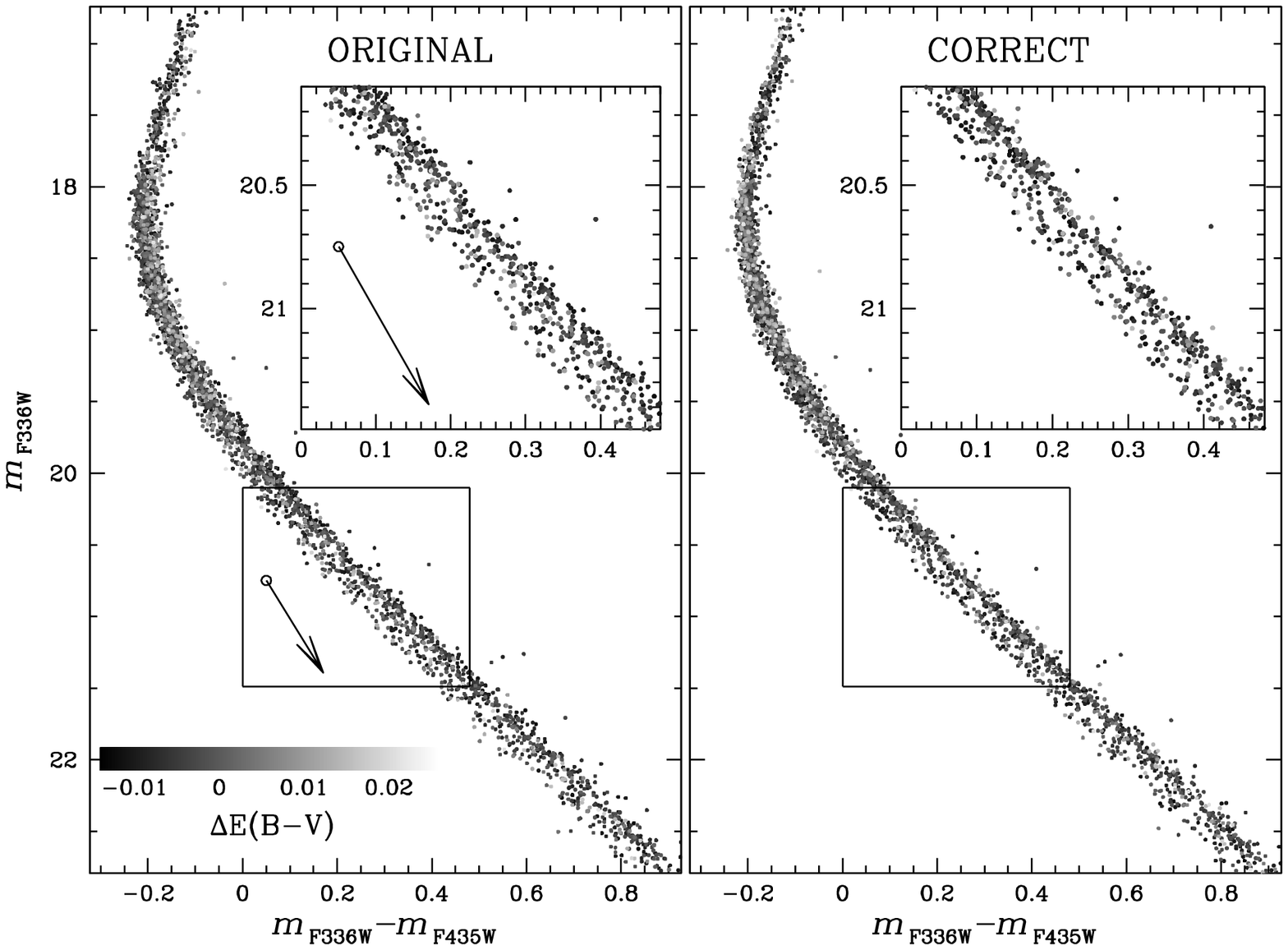}
\caption{Comparison of the original  (left panel) and the 
 differential reddening corrected (right panel)  $m_{\rm  F336W}$ vs.\ 
     $m_{\rm F336W}-m_{\rm F435W}$ CMD. The insets show a zoom around 
     the MS while the arrows indicate the reddening direction. 
     The gray levels corresponds to different differential reddening 
     variation as indicated in the figure.  }  
         \label{M3a}
   \end{figure}

 We find that the reddening variations are typically smaller 
     than $\Delta$E(B-V)$\sim$0.01 mag, and never exceed 0.026 mag.
     The reddening map we obtained is plotted in Fig.~\ref{M3b},
     where  we divide the field of view into 8 horizontal slices
     and 8 vertical slices, and plot $\Delta~E(B-V)$ as a function 
     of the Y (upper Panels) and X coordinate (right Panels).
     We have also divided the whole field of view into 32$\times$32 boxes
     and calculated the average $\Delta~E(B-V)$ within each of them.
     The resulting reddening map is shown in the lower-left Panel,
     where each box is represented as a gray square. The  levels of 
     gray are indicative of the amount of differential reddening as 
     shown in the upper-right plot.  As an example, a comparison of 
     the original $m_{\rm F336W}$ vs.\ $m_{\rm F336W}-m_{\rm F435W}$ CMD 
     is provided in Fig.~\ref{M3a}.

\section{The split Main Sequence}
\label{sec:MS}

\begin{figure*}[ht!]
\centering
\epsscale{.95}
\plotone{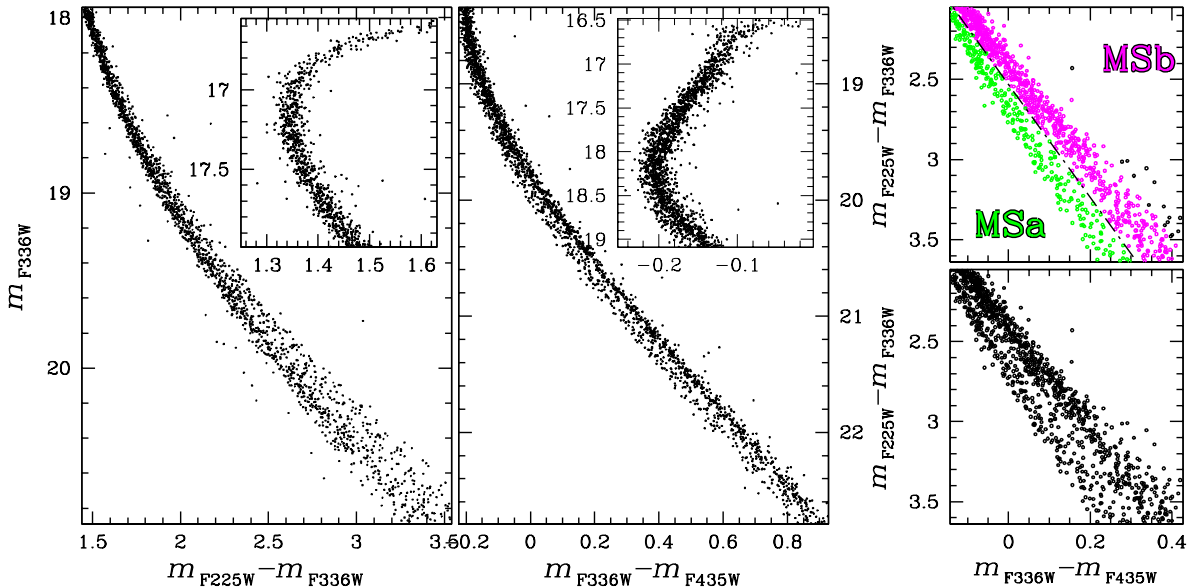}
\caption{ 
 \textit{Left panels:} 
         Proper-motion-selected, differential-reddening-corrected
         ${\it m}_{\rm F336W}$ vs.\ ${\it m}_{\rm F225W}-{\it m}_{\rm F336W}$ 
         (\textit{left panel}) and
         ${\it m}_{\rm F336W}$ vs.\ ${\it m}_{\rm F336W}-{\it m}_{\rm F435W}$  
         (\textit{middle panel}) CMDs for MS stars in NGC 6397. 
         The insets show a zoom of the SGB region.
 \textit{Right panels:}  
         the ${\it m}_{\rm F225W}-{\it m}_{\rm F336W}$ 
            vs.\ ${\it m}_{\rm F336W}-{\it m}_{\rm F435W}$ two-color
         diagram for MS stars with 19.2$<{m}_{\rm F336W}<$20.6. 
         The dot-dashed line in the upper panel was drawn by hand to
         separate the MSa and MSb stars, plotted green and magenta, 
         respectively.} 
         \label{CMDs}
   \end{figure*}

Figure~\ref{CMDs} shows CMDs with two 
 different color baselines: 
${\it m}_{\rm F336W}$ vs.\ ${\it m}_{\rm  F225W}-{\it m}_{\rm F336W}$  and 
${\it m}_{\rm F336W}$ vs.\ ${\it m}_{\rm  F336W}-{\it m}_{\rm F435W}$.
It is clear that the MS of NGC 6397 is split into two distinct sequences
 in both diagrams. 
  It is interesting to note that in the 
      ${\it m}_{\rm  F225W}-{\it m}_{\rm F336W}$ color, the majority of
      the stars are on the blue MS branch (hereafter MSb), while in the
      ${\it m}_{\rm F336W}-{\it m}_{\rm F435W}$ color system, the majority
      of stars are on the redder sequence.  We will explore and quantify
      this effect below.

The presence of a split MS is even more evident from the two-color
diagram shown in the right panels.  We have separated the two sequences 
into a MSa (green colors in the right upper panel of Fig.~\ref{CMDs}), 
and a MSb (magenta colors).

The insets of left and middle panel of Fig.~\ref{CMDs} show that 
the two MSs appear to merge close to the turn-off, and that the 
turn-off/subgiant-branch part of the CMD is narrow and well defined, 
implying that any age difference between the two populations
must be less than a few 100 Myrs.  Putting a better upper limit
on the possible age spread will require a measurement of the 
 overall C+N+O content for the stars in the two populations.
We note that the narrow magnitude distribution of SGB stars rules 
out the possibility that the double MS could be due to residual spatial
variations of reddening that are beyond the  sensitivity  of the
method that we used to correct them, as the reddening line is nearly 
perpendicular to the SGB sequence.  

There is additional evidence that the MS splitting is real.  First 
of all, the two MSs are visible everywhere in the field.   Furthermore, 
reddening would move stars in the same direction in any CMDs 
(e.\ g.\ higher reddening would move stars towards redder colors and 
fainter magnitudes, at odds with the observed behaviour of the MSa 
and MSb, which invert their relative position in different CMDs of 
Fig.~\ref{CMDs}.)


\begin{figure*}[ht!]
\centering
\epsscale{.65}
\plotone{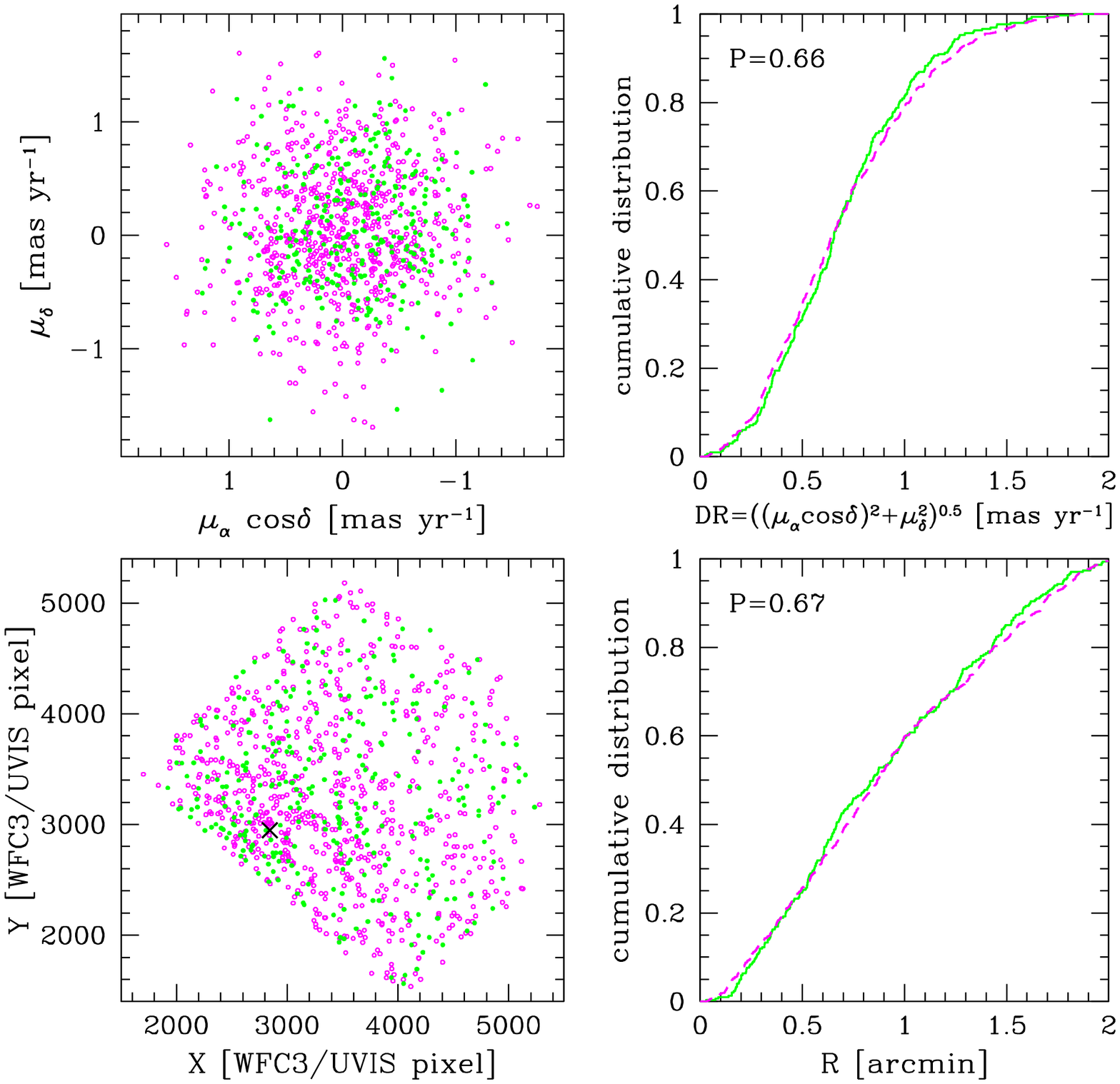}
\caption{{\textit Upper panels:} Distribution of MSa and MSb stars in
  the vector-point diagram of proper motions (right). The cumulative
  distributions is plotted in the left panel. {\textit Lower panels:}
  Spatial distributions (left) and cumulative radial distributions of
  MSa and MSb stars.} 
         \label{M5}
   \end{figure*}

 The upper-left panel of Figure~\ref{M5} shows the distribution 
     of MSa and MSb stars in the proper motion vector-point diagram, 
     with the colors that were assigned in Fig.~\ref{CMDs}.  The two
     samples show no difference in their proper-motion distributions.  
     To more quantitatively test this, in the upper-right panel we show 
     the cumulative distribution of proper motions. A Kolmogorov-Smirnov 
     test shows a vey high probability (P=66\%) for the null hypothesis 
     that the two populations have a similar proper-motion distributions
     The spatial distribution is plotted in the lower-left panel. 
     In this case, as well, the Kolmogorov-Smirnov test indicates that 
     there is no evidence for difference in the radial distributions 
     (P=67\%).


\begin{figure}[ht!]
\centering
\epsscale{.75}
\plotone{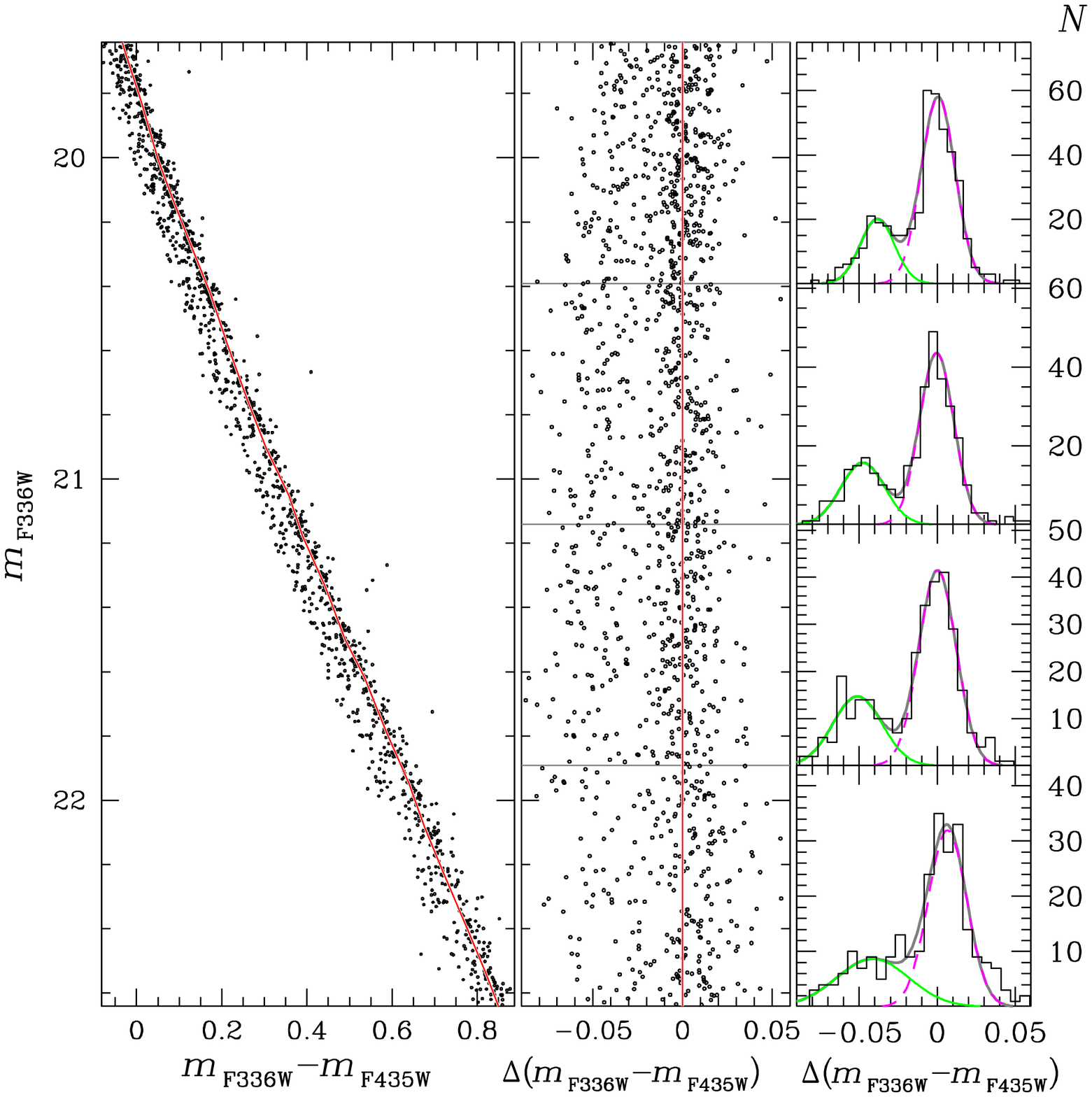}
   \caption{
       \textit{Left panel:} 
          A zoom of the region  of the ${\it m}_{\rm F336W}$ vs.\ 
          ${\it m}_{\rm F336W}-{\it m}_{\rm F435W}$ CMD from 
          Fig.~\ref{CMDs},  where the MS split is more evident.  
          The continuous line is a fiducial line of the MS. 
       \textit{Middle panel:} 
          The same CMD, after subtraction of the color of the 
          fiducial line.
       \textit{Right panels:} 
          The color distribution of the points in the middle panel in 
          four F336W magnitude intervals.  The continuous grey lines are 
          fits by a sum of the two Gaussians.
       }
\label{MSratio}
\end{figure}

In order to estimate the fraction of stars in each MS we followed the 
procedure illustrated in Fig.~\ref{MSratio}, which has already been used in 
several previous papers (e.\ g.\ Piotto et al.\ 2007).  The left panel shows 
the same ${\it m}_{\rm  F336W}$  vs.\ 
${\it m}_{\rm  F336W}-{\it m}_{\rm   F435W}$ CMD of Fig.~\ref{CMDs}, zoomed 
around the upper MS region, where the split is most evident.  The red 
line is the fiducial ridge line of MSb.  To determine it, we started by 
selecting a sample of MSb stars by means of a hand-drawn first-guess 
ridge line, and choosing the stars within a limited color range about 
this line.  We then calculated the median color and magnitude of MSb 
stars in intervals of 0.2 magnitude in F336W, interpolated these median 
points with a spline, and did an iterated sigma-clipping of the 
straightened sequence.  To obtain the straightened MS of the middle 
panel, we subtracted from the color of each star the color of the fiducial 
sequence at the  F336W magnitude of the star.  The color distribution 
of the points in the middle panel were analysed in four magnitude bins 
over the interval 19.6$<m_{\rm F336W}<$20.6.  The distributions have 
two clear peaks, which we fit with two Gaussians (magenta for the MSb 
and green for the MSa).  From the areas under the Gaussians, 71$\pm$3\% 
of stars turns out to belong to the MSb, and $29 \pm 3$\% to the MSa.  
Within the statistical uncertainties these fractions are the same in 
all magnitude intervals.

\begin{figure}[ht!]
\centering
\epsscale{.75}
\plotone{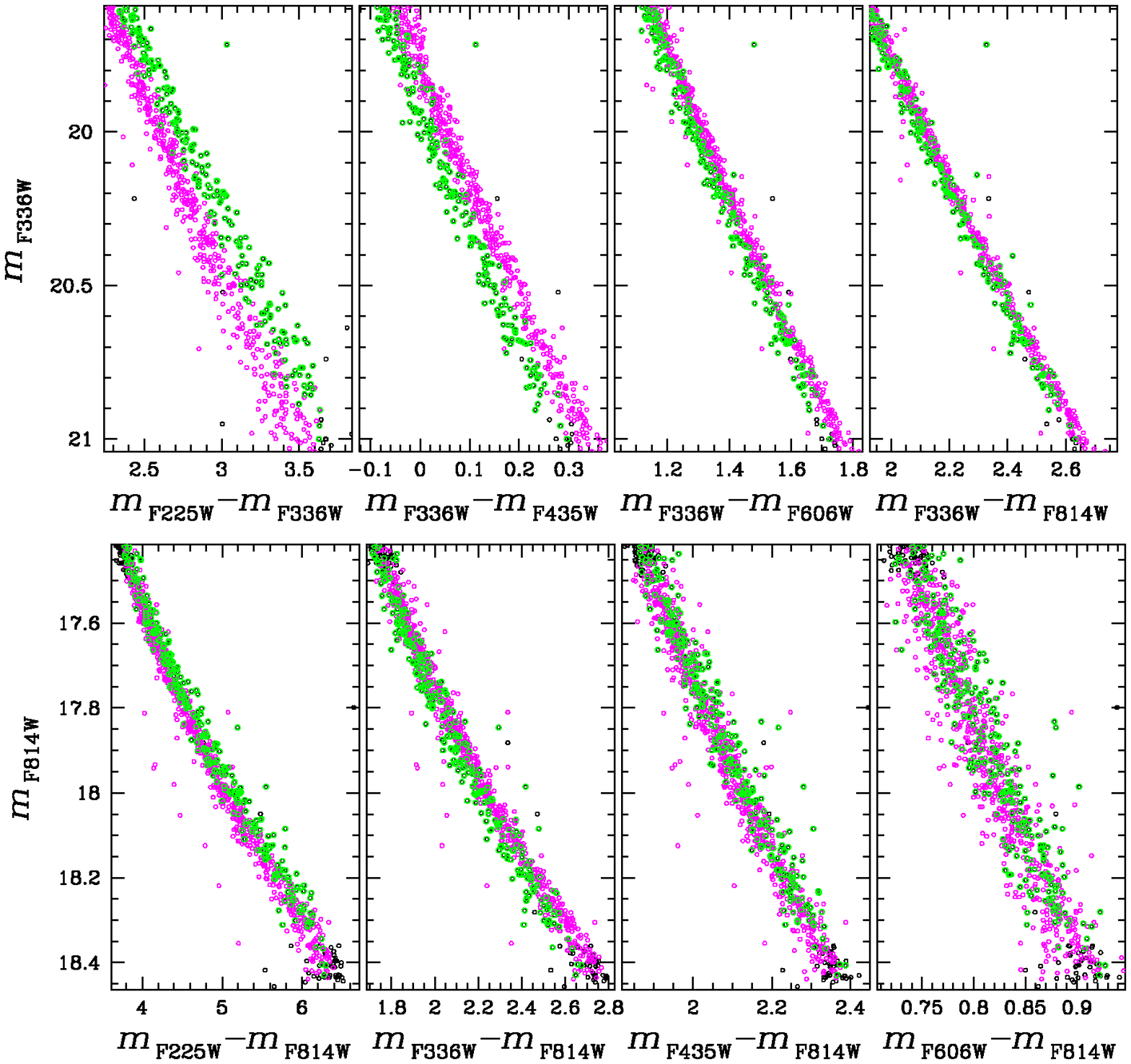}
\caption{
   \textit{ Upper panels:}   
       ${\it m}_{\rm F336W}$ vs.\ colors made with ${\it m}_{\rm F336W}$ 
       and F225W, F435W, F606W, and F814W for MSa stars (green) and 
       MSb stars (magenta).  MSa is bluer than MSb in all the colors 
       except ${\it m}_{\rm  F225W}-{\it m}_{\rm F336W}$.  
   \textit{ Lower panels:} 
      ${\it m}_{\rm  F814W}$ vs.\ colors made with ${\it m}_{\rm F814W}$
      and F225W, F336W, F435W, and F606W.  In these colors the MSa is 
      redder than the MSb with the exception of the CMD based on the 
      ${\it m}_{\rm F336W}-{\it m}_{\rm F814W}$ color.}  
\label{multiwaveA}
\end{figure}

Since we have photometry through five different $HST$ filters, we can 
study the relative location of the two MSs of NGC 6397 in a variety 
of CMDs constructed using all possible combinations of the photometric 
bands, as shown in Fig.~\ref{multiwaveA}.  The upper panels show the 
CMDs ${\it m}_{\rm F336W}$ vs.\ ${\it m}_{\rm X}-{\it m}_{\rm F336W}$ 
or ${\it m}_{\rm F336W}-{\it m}_{\rm X}$ (where X represents F225W, 
F336W, F435W, F606W, F814W).  We used the same color code to plot the 
MSa and MSb stars identified in the color-color diagram in Fig.~\ref{CMDs}.
The MSa has a ${\it m}_{\rm  F225W}-{\it  m}_{\rm F336W}$ color that is 
redder than the MSb, but the relative color is opposite of this in all 
the other ${\it m}_{\rm F336W}-{\it m}_{\rm X}$ colors. 

\begin{figure}[ht!]
\centering
\epsscale{.75}
\plotone{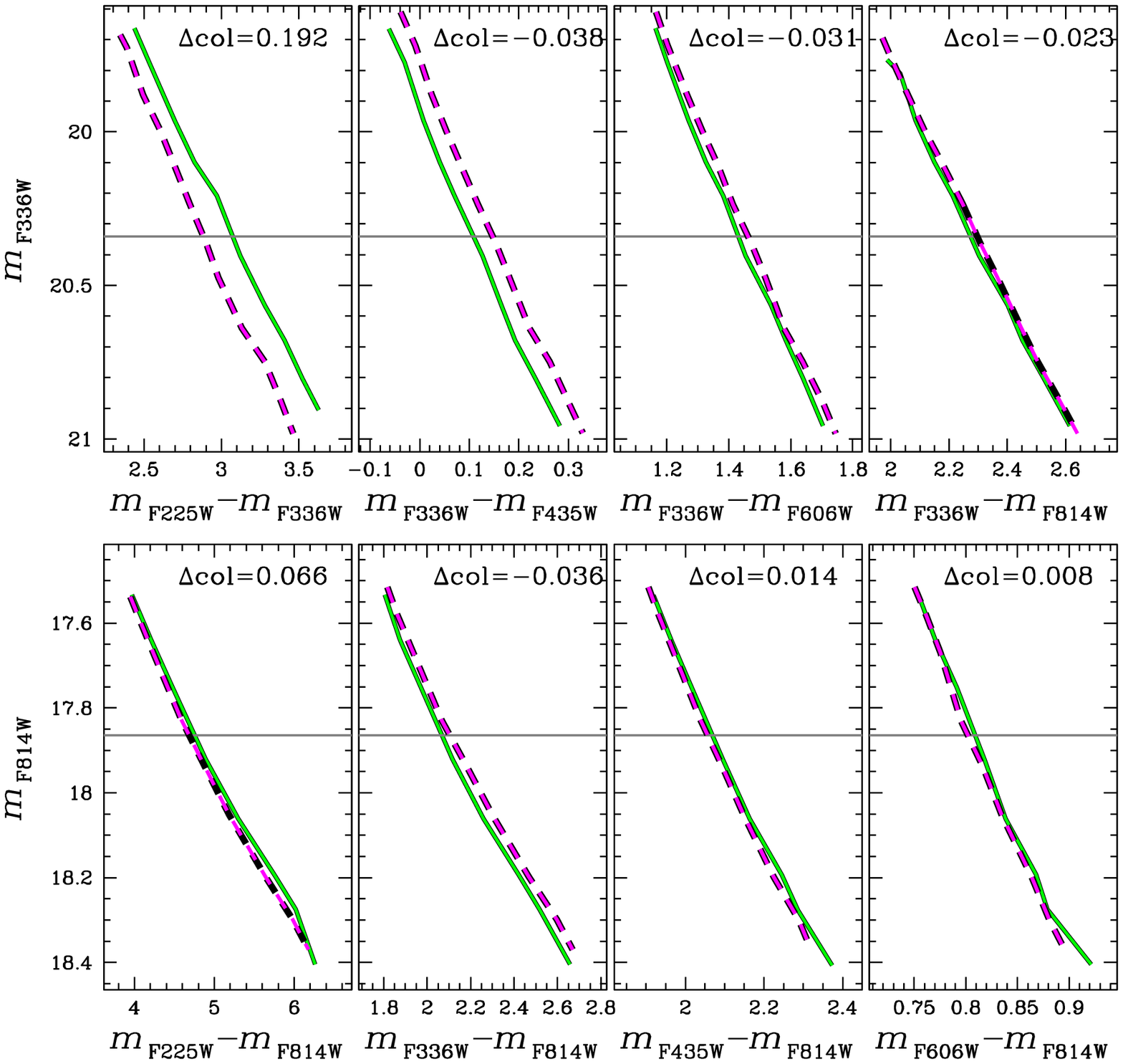}
\caption{
   \textit{ Upper panels:} 
       MS fiducials for the same CMDs shown in the upper panels of the 
       previous figure.  MSa is represented by the solid green line, 
       and MSb by the dashed magenta line.  We report at the top of 
       each panel the color separation of the fiducials of MSa and 
       MSb measured at the ${\it m}_{\rm F336W}=20.3$ level (indicated 
       by the horizontal grey line).  
   \textit{ Lower panels:} 
       MS fiducials for the CMDs in the lower panels of the previous
       figure.  The reported $\Delta(col)$ values are calculated at 
       ${\it m}_{\rm F814W}=17.9$.  }  
\label{multiwaveB}
\end{figure}

The lower panels of Fig.~\ref{multiwaveA} show the ${\it m}_{\rm F814W}$ 
vs.\ ${\it m}_{\rm X}-{\it m}_{\rm F814W}$ CMDs.  Here, the MSa is bluer 
than the MSb, with the exception of the already mentioned
${\it m}_{\rm F814W}$ vs.\ ${\it m}_{\rm F336W}-{\it m}_{\rm F814W}$
CMD.  The separation of the two sequences increases for larger color 
baselines in the remaining CMDs.
 To better visualize the relative positions of the two MSs, in
     Fig.~\ref{multiwaveB} we plotted the fiducials of the MSa and 
     the MSb derived for each CMD of Fig.~\ref{multiwaveA} following  
     the same procedure above described. 

%

%
\section{Discussion}
\label{sec:discussion}
In this paper
we have shown that the MS of NGC 6397 is  clearly
split into two distinct sequences.  Broadened or multiple MSs have 
been observed in other clusters, including  47 Tuc, $\omega$~Centauri, 
NGC 2808, and NGC 6752 (Anderson 1997, Bedin et al.\ 2004, 
Piotto et al.\ 2007).   In $\omega$~Centauri and NGC 2808 the multiple 
MSs have been associated with stellar populations that have large 
difference in their helium abundance, with the bluer MSs having a 
higher He-content than the redder ones (Norris 2004, Piotto et al.\ 2005, 
D'Antona et al.\ 2005).  Stellar evolutionary models predict that 
H-burning at high temperatures through the CNO cycle results in 
an enhanced production of He.  In addition to He, such models also 
predict enhanced production of N, and Na and depletion of C, and O.  
This pattern of enhancement/depletion has recently been confirmed 
among the blue and red MS stars in NGC 2808 by Bragaglia et al.\ (2010).

Since the seventies, spectroscopic studies of NGC 6397 giants have 
shown that the RGB stars exhibit a large spread in their C and N abundances 
(Bell et al.\ 1979). 
Significant star-to-star variations in the Na content have also been 
detected among unevolved TO stars (Gratton et al.\ 2001, 
Lind et al.\ 2009), thus demonstrating that such abundance variations
must have an intrinsic (primordial) rather than an evolutionary origin. 
NGC 6397 stars also show Na-O anticorrelation (Carretta et al.\ 2005, 2009,
L11).  Even more relevant for  the present discussion, 
L11 analysed the  chemical composition of a large
number of elements in 21 RGB stars of NGC 6397, and found that Na and O
abundances have a bimodal distribution, with about the 75\% of stars
being enriched in Na and depleted in O, while the remaining $\sim$25\% 
of the stars have a abundance patterns similar to field stars.  Similarly, 
Carretta et al.\ (2009) found that 30\% of the stars in their sample are 
Na-poor/O-rich.  L11 also demonstrated that the RGB of NGC 6397 is bimodal 
in the Str\"{o}mgren color index $c_{\rm  y}=c_{\rm 1}-(b-y)$, with the 
Na-poor and Na-rich stars populating the blue and red RGB sequences, 
respectively.  It is interesting to note that, in Sect.~3 we found 
that $71\pm3$\% of the MS stars populate the MSb.  This fraction is 
consistent with the 75\% of Na-rich, O-poor stars found by L11.  
Therefore, we suggest a connection between the MSb and the Na-rich, 
O-poor population of NGC 6397.  This would imply that the MSa represents 
the first generation, and its intermediate-mass stars polluted the cluster 
with gas rich in He, N, and Na  to make the second generation, MSb.

In an effort to quantify the observed MS split, we 
 have characterized the 
fiducial sequences by measuring the color difference between the MSa 
and MSb at the reference magnitude of $m_{\rm F814W}=17.9$ (Fig.~3).  
For an assumed apparent distance modulus of $(m - M)_{\rm F814W}$ = 12.2, 
this corresponds to an absolute magnitude of $M_{\rm F814W}$ =5.7.

To compare these observations against expectations from synthetic
photometry, we adopted the BaSTI isochrones (Pietrinferni et al.\ 2004) 
for the populations listed in Table~1, and determined $T_{\rm eff}$
and $\log g$ for MS stars at $M_{\rm F814W}=5.7$.  
We adopted the average C, N, and O abundance of N-poor stars as 
     measured by Carretta et al.\ (2005) for dwarf and SGB stars in 
     NGC 6397.

These temperatures and gravities (listed in Table~1) were then used 
to calculate model atmospheres with the ATLAS12 code (Kurucz 2005, 
Castelli 2005, Sbordone 2005), which allowed us to use specific chemical 
compositions.  We then used the SYNTHE code (Sbordone et al.\ 2007) 
to synthesize the spectrum from 1000\AA\ to 10000\AA\, and the resulting 
synthetic spectra were integrated over the transmission curve of each 
of our filters to produce the synthetic magnitudes and colors.  We did 
this separately for an MSa star, 
 using the composition listed in Table~1, and for an and MSb star, using the 
 three different composition options given in Table~1.  

\begin{figure}[ht!]
\centering
\epsscale{.75}
\plotone{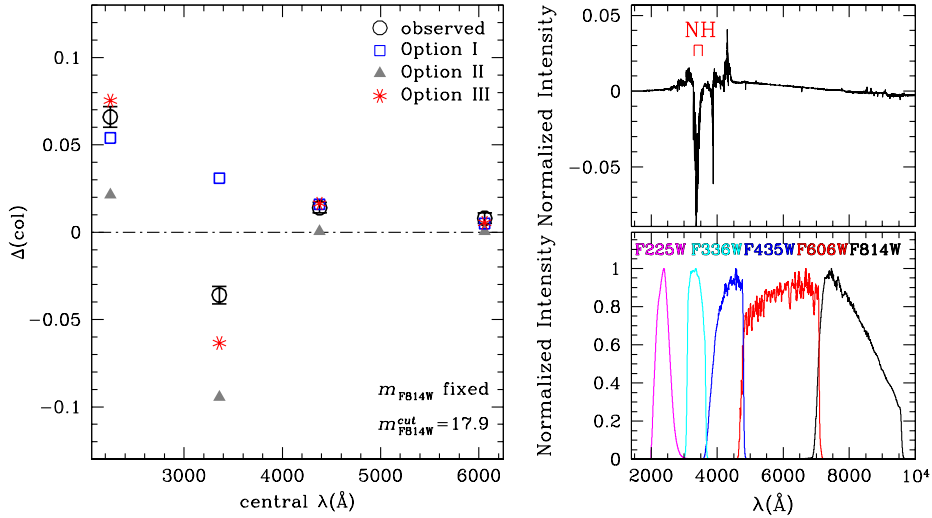}
\caption{
   \textit{Left panel:} 
       Color differences of the fiducial lines of MSa and MSb. The
       adopted colors are ${\it m}_{\rm X}-{\it m}_{\rm F814W}$, where
       X=F225W, F336W, F435W, and F606W. Each color difference is
       measured at ${\it m}_{\rm F814W}=17.9$, and is plotted as a 
       function of the central wavelength of the filter X.  Positive 
       or negative $\Delta$col values indicate that the MSa is bluer 
       or redder than the MSb.  
   \textit{
       Upper-right panel:} The difference between the synthetic spectrum 
       of a MSa and a MSb star (see text for more details).  
    \textit{Lower-right panel:} Normalized responses of the {\it HST} 
       filters used in this paper.  } 
\label{MODELS}
\end{figure}

The left-panel of Fig.~\ref{MODELS} shows a comparison of the observed 
$m_{\rm X}-m_{\rm F814W}$ color differences between MSa and MSb 
  against
the synthetic ones.  The blue squares indicate the color differences 
corresponding to Option I, where we assumed for the two MSs the same 
C, N, O mixture, but different He content.  There is a good agreement 
with the observed color differences in most bands, 
 but a signficant disagreement with the F336W band.
We conclude that Helium alone cannot account for the observed MS split. 

\begin{table}[ht!]
\begin{center}  
\scriptsize {
\caption{Parameters used to simulate synthetic spectra of an MSa and
an MSb star with $m_{\rm F814W}$=17.9, for the three assumed options.}
\begin{tabular}{ccccccc}
\hline
\hline
MS (Option) & $T_{\rm eff}$    & log g & $Y$    & [C/Fe] & [N/Fe] & [O/Fe]  \\
\hline
MSa (all) &     5373  & 3.26   &0.246   &  0.10  &-0.20   & 0.45 \\
MSb (I)   &     5398  & 3.26   &0.256   &  0.10  &-0.20   & 0.45 \\
MSb (II)  &     5373  & 3.26   &0.246   & -0.05  & 1.35   & 0.10 \\         
MSb (III) &     5398  & 3.26   &0.256   & -0.05  & 1.35   & 0.10 \\         
\hline
\hline
\label{paramMSs}
\end{tabular}
}
\end{center}
\end{table}

It is worth noting that Nitrogen mainly affects the F336W 
 photometry via the NH band around $\lambda \sim 3400$ \AA\ (e.\ g.\ 
Marino et al.\ 2008).  In Option II, we assume that the MSb stars in  
NGC 6397 are N-enhanced, but have the same He content.  We assumed 
for the MSb the average C, N, and O abundance found by Carretta 
et al.\ (2005) for N-rich stars.  

The colors that result from Option II are plotted as gray triangles in
Fig.~\ref{MODELS}.  There remains a significant discrepancy between 
the simulated and the observed color differences in the F336W and 
F225W bands, now in the opposite sense to that seen in Option I. 
Turning instead to Option III, with differences in both helium and 
the CNO elements, we see that the red asterisks in Fig.~\ref{MODELS} 
are in better agreement with all of the observed color differences,
 yet it is still not perfect for the F336W filter. 

A careful fine-tuning of the composition differences between the two 
MSs might remove the residual discrepancies between observed and theoretical 
points, but it is unclear whether that would tell us what is really going 
on without high-resolution spectra.   The difference between the synthetic 
spectra of a MSa and a MSb stars as calculated for Option III is shown 
in the upper-right panel of Fig.~\ref{MODELS}, while the band-passes of 
our filters are plotted in the bottom-right panel.

%

In conclusion, the observed color differences between the two MSs 
are consistent with the presence of two populations with different 
helium and light-element content.  Specifically, the MSa would correspond 
to the first stellar generation with primordial He, and O-rich/N-poor 
stars, while the MSb would  be made of stars enriched in He and N but 
depleted in O.  While the bluer MSs of $\omega$~Centauri and NGC 2808 
imply extremely high He abundances (up to Y$\sim$0.38), the MS of NGC 6397 
is consistent with a significantly smaller helium enhancement of 
second-generation stars at a level of $\Delta Y$=0.01, as already 
suggested by Di Criscienzo et al.\ (2010).  Note that we reached similar 
conclusions, based on the same technique described above, for 47 Tuc 
(Milone et al.\ 2011a) where we found a $\Delta Y$$\sim$0.02.

Finally, we recall that light-element variations have been detected in 
all GCs studied to date (see Gratton et al.\ 2004 for a review).  If the
interpretation of the split MS of NGC 6397 given above is correct, then 
either multiple or spread MSs should be observed in most (all?) GCs.  
In this paper (and in a similar paper on 47 Tuc), we have demonstrated 
that the combination of UV and visual photometry, 
 such as ${\rm m}_{\rm F225W}$, ${\rm m}_{\rm F336W}$, and ${\rm m}_{\rm  F435W}$, 
is a powerful tool to isolate stellar populations in the color-magnitude 
and color-color diagrams, allowing us to infer their He and CNO content.


\begin{acknowledgements}
G.P. and S.C. acknowledge partial support by  PRIN INAF 'Formation and Early
Evolution of Massive Star Clusters' and by ASI under the program ASI-INAF I/016/07/0. G.P. acknowledge partial support by the Universita' di Padova with the Progetto di Ateneo "Multiple Stellar Populations in Star Clusters".  J.A. acknowledges the support of STScI grant GO-11233.
\end{acknowledgements}


\bibliographystyle{aa}

\end{document}